\documentclass[aps,reprint,superscriptaddress]{revtex4-2}
\usepackage{graphicx}
\usepackage{dcolumn}
\usepackage{bm}
\usepackage[utf8]{inputenc}
\usepackage[T1]{fontenc}
\usepackage{booktabs, array, mathptmx, float, tabularx, booktabs, lipsum, multirow, amstext}
\usepackage{amsmath}
\usepackage{mathptmx}
\usepackage{mathrsfs}
\usepackage{upgreek}
\DeclareMathAlphabet{\mathcal}{OMS}{cmsy}{m}{n}
\usepackage{siunitx, xcolor}
\usepackage[version=4]{mhchem}

\usepackage{xr}
\begin{document}

\title{Dielectric permittivity extraction of MoS$_2$ nanoribbons using THz nanoscopy}
\author{Henrik B. Lassen}
\affiliation{Department of Electrical and Photonics Engineering, Technical University of Denmark, DK-2800 Kongens Lyngby, Denmark}
\affiliation{Current address: Department of Physics, Kyoto University, Kitashirakawa-Oiwake, Sakyo, Kyoto, Japan}
\author{William V. Carstensen}
\affiliation{Department of Electrical and Photonics Engineering, Technical University of Denmark, DK-2800 Kongens Lyngby, Denmark}
\author{Denys I. Miakota}
\affiliation{Department of Electrical and Photonics Engineering, Technical University of Denmark, DK-2800 Kongens Lyngby, Denmark}
\author{Ganesh~Ghimire}
\affiliation{Department of Electrical and Photonics Engineering, Technical University of Denmark, DK-2800 Kongens Lyngby, Denmark}
\author{Stela Canulescu}
\affiliation{Department of Electrical and Photonics Engineering, Technical University of Denmark, DK-2800 Kongens Lyngby, Denmark}
\author{Peter U. Jepsen}
\affiliation{Department of Electrical and Photonics Engineering, Technical University of Denmark, DK-2800 Kongens Lyngby, Denmark}
\author{Edmund J. R. Kelleher}
\email{Email: edkel@dtu.dk}
\affiliation{Department of Electrical and Photonics Engineering, Technical University of Denmark, DK-2800 Kongens Lyngby, Denmark}

\begin{abstract}
\noindent The nanoscale optical properties of high-quality MoS$_2$ nanoribbons are investigated using THz nanoscopy based on a scattering-type scanning probe.
The nanoribbons comprise a multi-layer core, surrounded by monolayer edges. 
A featureless complex permittivity spectrum covering the range 0.6-1.6~THz is extracted from experimental time-domain measurements through a minimization procedure, adopting an extended finite-dipole model of the probe-sample interaction.
Real-space mapping of the nanoribbon reveals variations in the local permittivity down to the instrument-limited resolution, on the order of 30~nm. 
Clustering analysis statistically identifies regions of lower apparent permittivity that we attribute to a high curvature at the edges of the nanoribbon causing an increase in local material strain or cross-talk in the measured signal with topography-induced measurement artifacts. 
The core of the nanoribbon contains two regions that follow tightly distributed, but slightly shifted Gaussian statistics in complex permittivity space, with the real part mean of both distributions lying around 5.4 and compatible with literature values of the static permittivity of thin-film MoS$_2$ reported previously. 
Our results show that the nanoribbons exhibit a modest degree of dielectric variation at the nanoscale that could be explained by heterogeneous doping or variations in the local defect density. 
We believe that our approach could be useful for the direct real-space measurement of dielectric disorder in other low-dimensional semiconducting material systems. 
\end{abstract}
\maketitle
\section{Introduction} 
The dielectric permittivity of a material is a measure of its response to an electric field and is a frequency-dependent complex quantity.
It is therefore a fundamental property of the material, governed by a specific chemical and structural composition, required in the design of semiconductor electronics.
Recently, dielectric disorder in nanoscale systems---fluctuations in local complex permittivity on optical length scales---has been identified to contribute strongly to variations in the optoelectronic performance and transport properties of two-dimensional (2D) materials~\cite{raja2019dielectric}.     
Transition-metal dichalcogenides (TMDs) are a family of 2D layered crystals, a subset of which are semiconducting and well-known to possess distinct properties in their mono- and few-layer form, namely a transition from indirect to direct electronic bandgap, compared to bulk. 
This behaviour marks out several TMDs in particular, including $\mathrm{MoS_2}$, $\mathrm{WSe_2}$, $\mathrm{MoSe_2}$, and more, as attractive materials for photonic and optoelectronic device applications, such as light-emitting diodes, photodetectors~\cite{Varghese2020}, and photovoltaics~\cite{Furchi2014}, among others, due to an increased efficiency of light-matter interaction driving a stronger photoluminescence and greater light absorption.
A further lowering of the physical dimensions can dramatically influence the intrinsic material properties~\cite{HuJAP2014, LiNaFNR2018, LiMos2NR2018}, boosting photoresponsivity and nonlinear effects~\cite{Chowdhury2020, Ghimire2023MoS2NR}.
The potential of TMD-based nanostructures, including MoS$_2$ nanoribbons, has pushed the development of numerous TMD synthesis methods forward, and emphasized the importance of dedicated characterization of TMD properties~\cite{Zhang2021, NatCom2015ALDMoS2, NLMoS2CVDgen2014, SHG2015WSe2opticsNat, Morozov2015OpticalConstantsDynamic}.

Angle-dependent Raman~\cite{7.5strain, BandGapEng_2013} and photoluminescence spectroscopies~\cite{Young2017, Zhang2015} have been widely applied in the study of layered materials, providing rich information on crystallographic structure including anisotropy, vibrational modes, strain, bandstructure, and defects; neither technique, however, directly captures the dielectric function and both are typically restricted in their ability to resolve features in materials or heterogeneities in optical properties smaller than is allowed by the diffraction-limited focusing of visible light (typically several hundreds of nanometers at best). 
Ellipsometry~\cite{Ermolaev2020} and impedance spectroscopy~\cite{zhao2017probing}, on the other hand, can directly record the frequency-dependent dielectric permittivity (albeit in very distinct frequency regions, either in the visible/near-infrared or radio frequency range), but are again subject to either the diffraction limit of optical systems or determine only an average macroscopic material response.
Similarly, time-domain methods in the infrared and low-frequency terahertz (THz) region of the electromagnetic spectrum, notably THz time-domain spectroscopy (TDS)~\cite{Jepsen2011}, have proven effective for material characterization because THz radiation, in particular, is commensurate with the energy scale of free carriers in materials leading to a strong interaction.
THz-TDS has been widely applied to semiconducting and metallic thin films, including graphene, to investigate complex permittivity and conductivity, without contacts~\cite{buron2012graphene, Yan2015}.   
However, the challenge is again the spatial resolution, which is now limited to several tens or even hundreds of micrometers because of the low-frequency (typically 0.3-3~THz) of the probing radiation.  

Scanning probe microscopy (SPM) techniques, such as atomic force microscopy (AFM) using sharpened (conductive) tips with an apex radius on the order of several tens of nanometers, can be used to explore material properties with significantly enhanced spatial resolution and are therefore suitable for studying nanostructured and heterogeneous materials locally.
Electrostatic force microscopy (EFM)---an AFM-based technique exploiting the change in capacitance between a voltage-biased probe and the sample surface to infer dielectric properties---is one example that has been applied to thin-film TMDs for recovery of the local electrostatic dielectric constant at (or very close to) DC~\cite{Hou2022QuantificationDielectricConstant}, but importantly does not provide information of dielectric relaxation behavior in even a limited frequency range.
Scattering-type scanning near-field optical microscopy (s-SNOM)~\cite{Knoll2000, Keilmann2004}, essentially a modified AFM with an external light source, equipped with infrared or THz illumination has become a powerful technique for nanoscale imaging and spectroscopy in a technologically relevant frequency range, combining the benefits of both sensitivity to free-carrier absorption of far-field THz-TDS, with the nanoscale resolution of other surface probes. 
The spatial resolution of s-SNOM is essentially agnostic to the wavelength of the illuminating light source~\cite{Chen2019ModernScatteringTypeScanning}, and is rather defined by the radius of the scanning probe (or tip) at its apex. THz-SNOM, therefore, probes a similar volume in real-space compared to SNOM illuminated with visible light. 
Significantly, however, the mismatch between the free-space wave-vector and the (in-plane) momenta of tip-scattered light at THz frequencies can reach $10^3$, meaning THz-SNOM probes deeply into the near-field regime.      
When the light source is a broadband THz pulse covering a wide frequency range, THz-enabled SNOM becomes a potent tool for nanoscopy of advanced materials. 
The challenge of SPM methods is often to accurately represent the tip-sample interaction. 
Quantitative extraction of material properties from experimental near-field scattering data in s-SNOM is often a formidable task, but several attempts in the recent literature have successfully applied an underlying model of the physical tip-sample system that can be inverted to relate fundamental parameters to measurable quantities~\cite{Govyadinov2013,Govyadinov2014, McLeod2014, Ritchie2022,zizlsperger2024situ}, without needing to make model assumptions about the dielectric function, and in the case of nanoscopy its spectral dependence.  
The inversion can be performed with an analytic approximation~\cite{Govyadinov2013} or using an iterative numerical minimization algorithm~\cite{mooshammer2018nanoscale}, with both demonstrated to yield robust and reliable output.     
Here, we use ultrafast THz pulses with a useful bandwidth spanning 0.6-1.6~THz for nanoscopy in an s-SNOM setup to interrogate MoS$_2$ nanoribbons and recover their complex dielectric response using a numerical minimization procedure based on an extended finite-dipole model for a layered material system~\cite{Cvitkovic2007, Hauer2012, Wirth2021}. 
Subsequent nanoscale THz imaging of the nanoribbon and a clustering analysis of the spatially-dependent dielectric data is used to identify regions strongly influenced by edge-effects where high surface curvature could indicate an impact from local strain or cross-talk from topographic artifacts, together with regions in the core of the nanoribbon where we observe two clearly distinguishable areas defined by tightly distributed Gaussian statistics that we believe capture directly nanoscale variations in the materials dielectric response.

\section{Materials and methods}
\subsection{Nanoribbon growth}

\begin{figure}[!ht]
    \centering
    \includegraphics[width=0.9\linewidth]{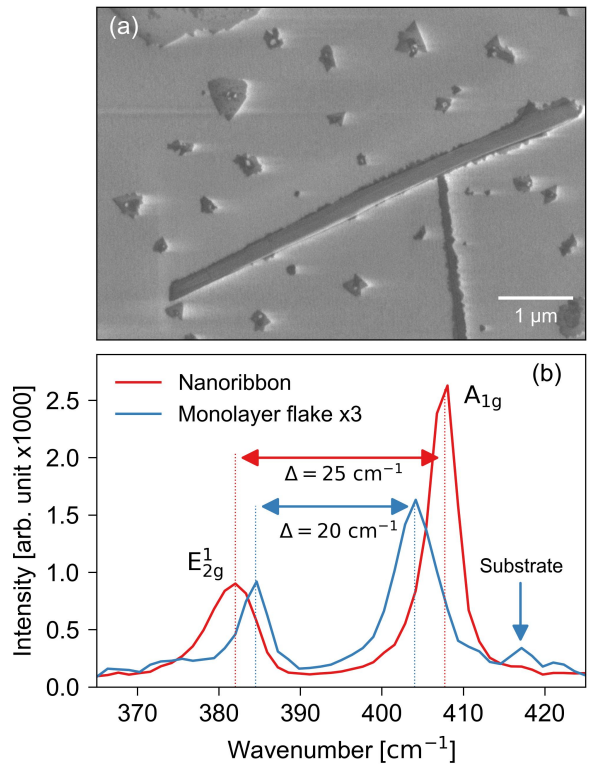}
    \caption{(a) SEM image of a characteristic $\mathrm{MoS_2}$ nanoribbon surrounded by triangular 2D and 3D crystallites of $\mathrm{MoS_2}$, and (b) Raman spectrum obtained from a region of the $\mathrm{MoS_2}$ nanoribbon (blue) and on a 2D $\mathrm{MoS_2}$ crystal (orange). 
    The signal from monolayer $\mathrm{MoS_2}$ has been multiplied by a factor of 3 for better representation.}
    \label{fig:raman}
\end{figure}
The $\mathrm{MoS_2}$ nanoribbons were grown on a \emph{c}-cut sapphire substrate.
Synthesis was a two-step process in which ultra-thin oxide films of $\mathrm{MoO_3}$-x (with many oxygen vacancies) grown by pulsed laser deposition were sulfurized at the second step in the presence of a NaF layer, the details of the process can be found in our previous works~\cite{Miakota2023MoS2NR, Ghimire2023MoS2NR}. 
The technique is similar to that described by Li~\emph{et al.} in Ref.~\cite{LiNaFNR2018} and shares many similar ideas on liquid phase creation and vapor-liquid-solid phase reaction.
Briefly, the growth process evolves via the formation of the Na–Mo–O liquid phase, which mediates the formation of $\mathrm{MoS_2}$ multilayer nanoribbons in a sulfur-rich environment~\cite{LiNaFNR2018, Miakota2023MoS2NR, Ghimire2023MoS2NR}. 
The nanoribbons crystallize predominantly in the 2H stacking orientation~\cite{Ghimire2023MoS2NR}.  
Due to strong in-plane covalent bonding and weak out-of-plane van der Waals (vdW) interactions, layered 2D materials possess a strongly anisotropic dielectric tensor~\cite{Ermolaev2021GiantOpticalAnisotropy}, with distinct in-plane and out-of-plane components.   
Figure~\ref{fig:raman}a shows a secondary-electron scanning electron microscope (SEM) image of a typical $\mathrm{MoS_2}$ nanoribbon on sapphire with a length of 10~$\mu$m and a width of less than 0.5~$\mu$m, resulting in a length-to-width ratio of nearly 20. 
Isolated 2D and 3D crystallites of MoS$_2$ can also been seen in the SEM image. 
A typical Raman spectrum of such samples taken using a laser with a wavelength of 532~nm is shown in Fig.~\ref{fig:raman}b. 
The spectrum shows two major characteristic Raman peaks of $\mathrm{MoS_2}$ arising from the in-plane $E^1_{2g}$ and out-of-plane $A_{1g}$ Raman modes. 
The peak position difference between the Raman modes of 25~cm$^{-1}$ usually denotes a bulk $\mathrm{MoS_2}$ response, which is in a good agreement with previous reports~\cite{AnomalMoS2vibrLee2010}. 
It should be noted that a strong $E^1_{2g}$ Raman peak shift can be experienced for the as-grown samples due to the strain presence in as-grown $\mathrm{MoS_2}$ nanoribbons. 
Here, we exclude the presence of residual strain in the $\mathrm{MoS_2}$ nanoribbon, because for the $\mathrm{MoS_2}$ nanoribbons grown, the strain can be released via rupture of the nanoribbons or the folds of the constituent $\mathrm{MoS_2}$ nanoribbon top layers~\cite{Miakota2023MoS2NR}, which can be seen in Fig.~\ref{fig:raman}a, further supported by analysis of the AFM topography map discussed later (see Supplementary Information), where the core of the nanoribbon shows a very small surface curvature that has been associated with low local strain~\cite{rahaman2017highly}.

\begin{figure*}[!t]
    \centering
    \includegraphics[width=0.9\linewidth]{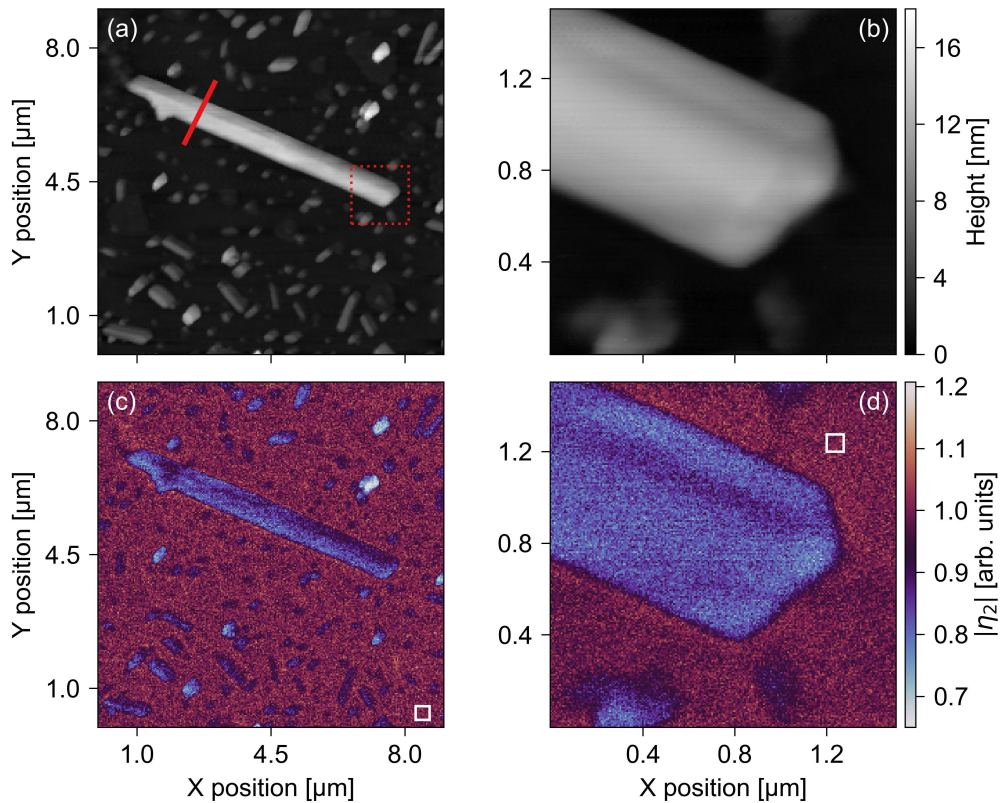}
    \caption{(a) AFM images of the selected $\mathrm{MoS_2}$ nanoribbon and (b) a zoom-in scan of one end of the nanoribbon as indicated by the red dotted square in (a). 
    The red line in (a) indicates the position of the spectroscopic line scan (THz nanoscopy). 
    (c) and (d) show the corresponding white-light mode imaging maps of the near-field contrast (second-order demodulated scattered signal). 
    The white squares in (c) and (d) indicate the areas used to generate a mean substrate response for normalization.}
    \label{fig:sample_characteristics}
\end{figure*}

The exact dimensions of the $\mathrm{MoS_2}$ nanoribbon used in this study can be inferred from the AFM images in Fig.~\ref{fig:sample_characteristics}a and Fig.~\ref{fig:sample_characteristics}b, showing an overview and zoom-in (corresponding to the area indicated by the dotted red box in Fig.~\ref{fig:sample_characteristics}a). 
The thickness of the nanoribbon (as confirmed by our AFM measurements, \emph{e.g.},~Fig.~\ref{fig:sample_characteristics}a) varies between approximately 10~nm and 15~nm, correspondingly the multilayer core of the nanoribbon consists of >15 layers suggesting the properties approach that of the bulk crystal. 
Notably, monolayer edges of the $\mathrm{MoS_2}$ nanoribbon that can be observed in the SEM imaging are not visible with AFM. 
It should be noted that the SEM and AFM were not performed on the same nanoribbon. 
The reason for discrepancies between the SEM and AFM images can be twofold: i) these edges are not pronounced in the selected nanostructure or ii) a limitation of the instrument: compared to SEM, AFM has a poor depth of field, limiting its ability to resolve features with large height differences. 
The abrupt change in thickness between the multilayer nanoribbon (15L) and the adjacent monolayer (1L) could make the monolayer edge undetectable by AFM.

\subsection{Near-field imaging and nanoscopy}
The near-field microscope used is a commercial instrument (Attocube THz-NeaSCOPE) equipped with an integrated THz time-domain spectroscopy module (Attocube/Menlo Systems TeraSmart). 
THz pulses with a useful bandwidth covering the spectral range 0.6-1.6~THz are generated and coherently detected by a photoconductive antenna pair. 
A conductive AFM tip (with a shank length of 80~$\mu$m and typical average tip radius of <40~nm, Rocky Mountain Nanotechnology, 25PtIr200B-H), operated in non-contact tapping mode (at a nominal frequency of 80~kHz), acts as a nanofocussing surface probe, allowing for simultaneous capture of sample topography together with measurements of the scattered near-field signal in a single scan of the sample.
Measurements are performed in a nitrogen rich environment to minimize the presence of water-vapor absorption lines in the detected THz spectra. 
Background removal is accomplished by demodulation of the scattered field to recover the near-field signal that is most pure in higher overtones of the tapping frequency (typically orders $m$ = 2-4 are used for data retrieval due to diminishing signal-to-noise ratio affecting data quality at orders above 4)~\cite{Knoll2000}. 
The scattered THz waveform is detected in the time-domain in two distinct modes of detection that are common to other THz-SNOM systems. 
Firstly, white-light (WL) detection offers rapid imaging of the surface by resolving the electric field at the principal peak of the waveform (this corresponds to a spectrally integrated near-field response~\cite{jing2023phase}). 
In this case, the WL signal, $\hat{E}$, is associated with both the amplitude and phase of the scattered radiation. 
In contrast, TDS mode (or nanoscopy) records the full scattered THz waveform in time from which the amplitude and phase spectra can be recovered using Fourier transformation, as is widely established in far-field THz-TDS~\cite{Neu2018}.
The incident electric field is \emph{p}-polarized to maximize the scattered signal~\cite{schneider2005scattering} and focused to the tip-sample interface using reflective optics at an angle of 30 degrees to the surface normal.
Due to the axisymmetry and elongated geometry of the AFM probe, it is widely known that the tip is dominantly polarizable parallel to its long axis, resulting in a larger out-of-plane polarizability~\cite{Yao2021ProbingSubwavelengthInplane}.
Consequently, although the scattered signal measured due to the tip-sample dipole is affected by the in-plane and out-of-plane components of the dielectric permittivity tensor~\cite{ruta2020quantitative}, the large in-plane momentum of fields scattered by the probe makes the technique most sensitive to changes in the out-of-plane component.
Thus, in the discussion of our results we consider an effective dielectric permittivity, acknowledging that this can be a complex mix of diagonal tensor elements for a uniaxial anisotropic crystal. 
We note, however, that under certain conditions the effective permittivity is separable: by exploiting parametrization and information from specific partially screened substrate resonances the dielectric tensor of bulk 2H-WSe$_2$ was determined in a specific frequency range in the mid-infrared~\cite{ruta2020quantitative}. 
Other efforts to recover dielectric tensor elements from s-SNOM data have included the use of guided modes in the near-infrared~\cite{norgaard2024near} or specially adapted probes~\cite{yao2021probing}.

\subsection{Modeling and inversion of the scattering problem}
The scattering of a low-frequency electric field by the combined tip-sample system can be described by several self-consistent, quasi-electrostatic models of the problem~\cite{Knoll2000,Cvitkovic2007,McLeod2014}, together with extensions to allow for layered structures, including thin-films on a bulk substrate~\cite{Hauer2012,Govyadinov2014,Wirth2021}, based on a transfer matrix formalism~\cite{Zhan2013}.
The finite-dipole model (FDM)~\cite{Cvitkovic2007}, and its extensions for layered structures~\cite{Hauer2012,Wirth2021}, approximates the probe as prolate spheroid. 
The scattered field is determined by $E_\mathrm{sca}=\alpha_\mathrm{eff}(1+r_p)^2E_\mathrm{inc}$, where $\alpha_\mathrm{eff}$ is the effective polarizability of the tip-sample system, $E_\mathrm{inc}$ is the incident field, and $r_p$ is the far-field reflection coefficient that is often reasonably ignored as it varies slowly relative to the spatial length-scales of typical measurements. 
This assumption is especially valid at THz frequencies with a sub-mm spot size.
Within the framework of the FDM,
\begin{equation}
\alpha_\mathrm{eff} \propto 1 + \frac{1}{2}\frac{\beta f_0}{1 - \beta(\omega, q) f_1},
\label{eq:effective_polarizability}
\end{equation}
where $f_{0,1}$ represent geometric functions describing the characteristics of the tip (see the Supplementary Information and references therein for details) and $\beta$ is the complex near-field reflection coefficient that depends on frequency, $\omega$ and the in-plane momentum, $q$ of the electric near-field. 
Importantly, $\beta$ carries information about the material properties of the sample, including the dielectric permittivity, which in its simplest form is written as $\beta=(\epsilon-1)(\epsilon+1)$ (see Supplementary Information for details).  
Evidently then the process of extracting the material properties follows from an inversion of the scattering problem. 
This is made more complex by the fact that the height of the tip above the sample surface is not fixed; in order to remove background and isolate the scattered near-field the tip oscillates.
Demodulation involves a Fourier decomposition into a series of harmonic orders that describe the overall scattering during a full oscillation cycle. 
We thus use an algorithmic approach, similar to Refs.~\cite{mooshammer2018nanoscale,Ritchie2022}, where a numerical routine aims to minimize the deviation between measured data and a corresponding scattered signal calculated from the dielectric function using the layer-extended FDM model (see Supplementary Information for further details of the inversion algorithm). 

\section{Results and discussion} 
We first show WL imaging of a selected nanoribbon in Fig.~\ref{fig:sample_characteristics}c (overview: 9x9~$\upmu$m at 36~nm per pixel) and Fig.~\ref{fig:sample_characteristics}d (zoom-in: 2x2~$\upmu$m at 10~nm per pixel).
Here we choose to present the second-order demodulated scattered near-field contrast, $\eta_2=\hat{E}_2^{\mathrm{sample}}/\hat{E}_2^{\mathrm{ref}}$, which is normalized to an average substrate response (or reference) corresponding to the mean signal within the white solid boxes, respectively. 
Relative to the substrate, the MoS$_2$ nanoribbon, and its crystallites, show a significantly lower scattered near-field signal.
This indicates that the magnitude of the permittivity of MoS$_2$ regions is lower than that of \emph{c}-cut sapphire.
More notable, are clear variations in contrast along the nanoribbon and between the nanoribbon and certain crystallites, some of which show a strikingly depressed contrast.
A transition region between the substrate and the nanoribbon accents the edge. 
Line-cuts through this area of the zoomed WL image, approximately perpendicular to the edge, for demodulation orders 2-4 (see Fig.~\ref{fig:si_topographic_impact_on_optical}b) together with the corresponding line-cut through the AFM topography (see Fig.~\ref{fig:si_topographic_impact_on_optical}a) indicate the instruments spatial resolution to be approximately 50~nm.
The resolution is principally governed by the sharpness of the tip at its apex (typically 30~nm based on SEM imaging of multiple tips), but it is also known that manifest edge effects can influence the image contrast in s-SNOM, including abrupt changes in the surface topography~\cite{chen2021rapid} and the dielectric environment of two dissimilar materials (particularly pronounced at the interface of an insulator and a metal, giving rise to an asymmetric transition of the image contrast)~\cite{mastel2018understanding}.
Although we see no clear asymmetry in our transition region across the nanoribbon edge over all inspected orders of the demodulated signal (due to the relatively small change in the dielectric values of the MoS$_2$ and the sapphire substrate), we do note a small peak in the relative contrast in the fourth-order scattered signal before it drops moving onto the nanoribbon that can suggest the measurement in this region is weakly influenced by a topographic artifact~\cite{chen2021rapid}. 

\begin{figure}[!t]
    \centering
    \includegraphics[width=0.9\linewidth]{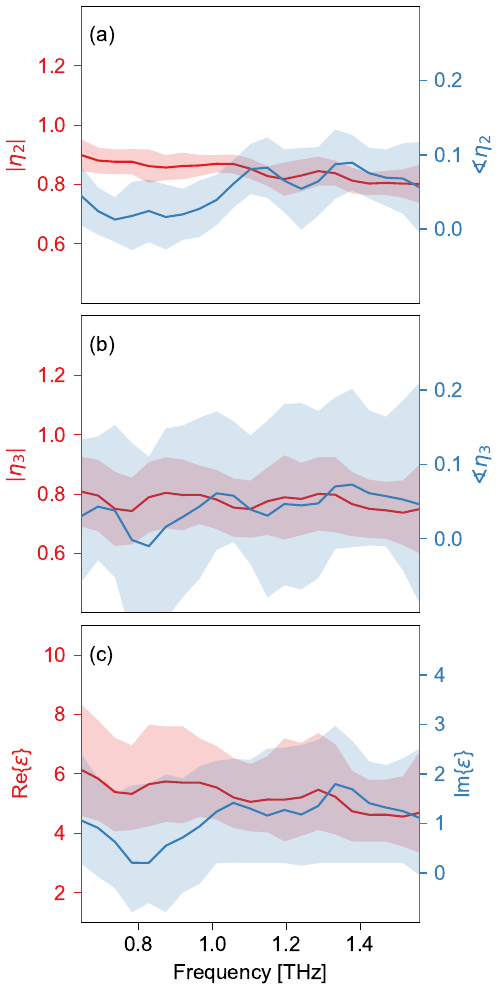}
    \caption{Spectrally-resolved THz nanoscopy; 
    (a) second-order contrast and (b) third-order contrast. 
    The solid lines indicate the response of 50 adjacent scans at the center of the nanoribbon and the filled area indicates the standard deviations. 
    (c) Extracted effective complex permittivity spectrum (see Supplementary Information for details of the algorithmic minimization procedure).}
    \label{fig:spectroscopy}
\end{figure}
Next we perform THz nanoscopy along a specific line, as indicated on Fig.~\ref{fig:sample_characteristics}a. 
The full scattered THz waveform is captured in the time-domain, with a spatial step-size of approximately 10~nm, starting and ending on the substrate and traversing the complete short axis of the nanoribbon. 
As before when WL imaging, the captured signal is demodulated and normalized to a reference comprising an average response of the substrate to obtain a relative contrast for a given harmonic order (see Supplementary Information for details regarding the normalization procedure).
The hyperspectral frequency-position data is spatially averaged over the central region of the nanoribbon where we observe little to no dependence in the contrast as a function of position. 
The resulting complex Fourier spectrum, and its corresponding standard deviation, for orders two and three is reported in Fig.~\ref{fig:spectroscopy}a and Fig.~\ref{fig:spectroscopy}b, respectively.
Both the real and imaginary parts of the contrast exhibit a rather featureless spectrum within the finite bandwidth limits of our probe. 
The strength of the contrast for the higher demodulation order is marginally greater, but this is accompanied by an increase in the standard deviation due to the corresponding reduced signal-to-noise ratio. 
The imaginary part is small and appears to be a near-constant value just above zero.

To extract the effective complex permittivity from the spectrally-dependent scattering contrast we utilize a numerical minimization procedure to invert the FDM (see the Supplementary Information for details of the algorithm used).
Due to the uniaxial anisotropy of the \emph{c}-cut sapphire substrate, its static effective permittivity was modeled by taking the geometric mean of the in-plane and out-of-plane components, $\varepsilon_{\mathrm{sub}_{||}}=9.46$ and $\varepsilon_{\mathrm{sub}_{\perp}}=11.68$~\cite{Grischkowsky1990}.    
The extracted permittivity shown in Fig.~\ref{fig:spectroscopy}c has a featureless spectrum, with only small ripples that we attribute to a residual error in the elimination of background contributions to the near-field signal arising from the instruments' response function and small-scale amplitude fluctuations or phase drift in the light source.
Although an additional normalization step by signal harmonics can provide effective suppression of such spectral artifacts at the cost of reduced material contrasts~\cite{mester2022high}, for the specific geometry and dimensions of our isolated nanoribbon normalization to the reference substrate alone yielded the best results.    
The flat spectral response is indicative of a low carrier density. 
Furthermore, the lowest frequency infrared-active phonons in single-layer and bulk MoS$_2$ reside at frequencies above 10~THz~\cite{molina2011phonons}, well beyond the bandwidth of our probe and therefore we do not expect to be sensitive to even the tails of these structural resonances. 
In fact, the real-part mean, $\mathrm{Re}\{\bar{\varepsilon}\}=5.43$, is in excellent agreement with recent measurements of the static permittivity of MoS$_2$ thin-films performed using EFM~\cite{Hou2022QuantificationDielectricConstant}. 

\begin{figure*}[!t]
    \centering
    \includegraphics[width=0.9\linewidth]{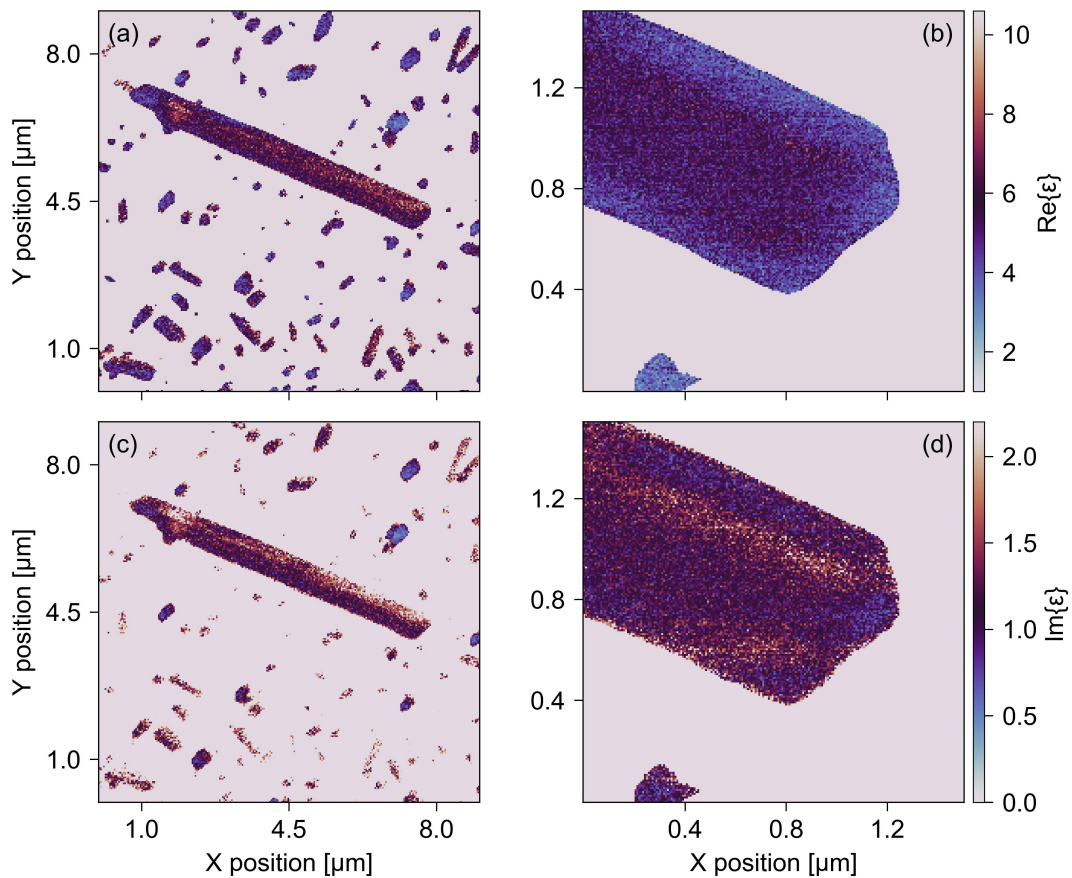}
    \caption{Spatially-resolved maps of the real (a,b) and imaginary (c,d) parts of the complex permittivity, extracted from WL imaging of the selected MoS$_2$ nanoribbon, overview (a,c) and zoom-in (b,d).}
    \label{fig:WL_epsilon}
\end{figure*}
The featureless spectrum, confirmed by our line-scan measurements using detection of the full scattered THz waveform (TDS-mode) to recover amplitude and phase information, allows us to further analyze the WL imaging---where detection recovers a weighted spectrally-integrated electric near-field response---in order to investigate nanoscale variations in the effective permittivity. 
Using the same inversion procedure as introduced above, but this time applied to the WL data (in Figs.~\ref{fig:sample_characteristics}c and d), we show the extracted real (a,b) and imaginary parts (c,d) of the effective complex permittivity in Figs.~\ref{fig:WL_epsilon}a-d for the overview (a,c) and zoom-in (b,d), respectively.
Throughout the inversion, we assume a constant value for the phase of the scattered field taken to be the frequency-resolved average from TDS measurements (see the Supplementary Information for details). 
While this assumption is imperfect, due to the nature of the WL detection being a mix of contrast due to changes in amplitude and phase, we confirm at several spatial locations across the nanoribbon that this approximation is only a small correction to the full response and variations in the measured contrast, leading to a spatial distribution of extracted permittivities, is dominated by contributions from changes in amplitude (see Fig.~\ref{fig:si_waveforms}).
In contrast to recording the full waveform at each spatial pixel, WL detection is relatively fast.                
Thus, we believe this approach allows us to rapidly determine nanoscale regions within the entire nanoribbon with distinct distributions of the permittivity that would otherwise be obscured using far-field optical probes that at best could perform a spatially averaged response over a small ensemble of nanoribbons or not captured with single-point THz nanoscopy.

\begin{figure*}[htb]
    \centering
    \includegraphics[width=0.9\linewidth]{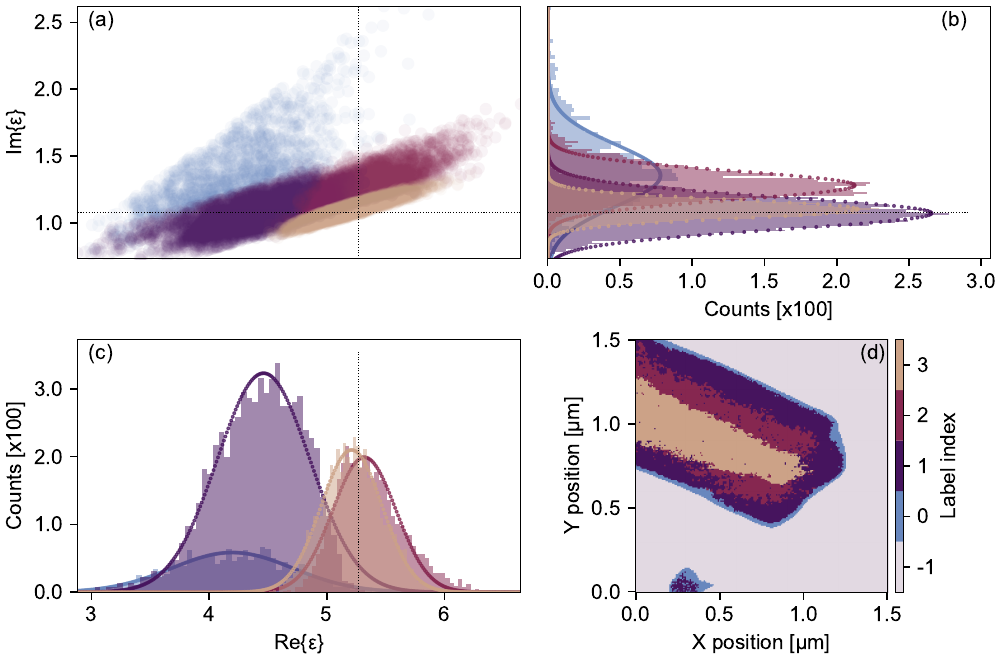}
    \caption{Clustering analysis of the spatially resolved permittivity data. (a) Four distinct clusters in the complex plane. Projections of the four dominant cluster distributions onto the imaginary (b) and real (c) axis. (d) Mapping of the clusters back to real-space to identify variations in the nanoscale permittivity associated with local changes in material properties.}
    \label{fig:epsilon_distribution}
\end{figure*}
We perform a clustering analysis of the real-space data in order to identify heterogeneities in the effective permittivity at the nanoscale. 
In Fig.~\ref{fig:epsilon_distribution}a we show a phase-space representation of the output of a Bayesian Gaussian mixture model that identifies four distinct clusters in the complex plane. 
In Fig.~\ref{fig:epsilon_distribution}b and c, these clusters are projected onto the real and imaginary axis and the resulting distributions are fitted with a Gaussian function, respectively. 
The horizontal and vertical black dotted lines represent the mean values of the real and imaginary parts, as determined from the THz nanoscopy data (Fig.~\ref{fig:spectroscopy}c).
The clusters (indexed 0-3, with -1 being the substrate) are then mapped back onto the real-space image of the nanoribbon (Fig.~\ref{fig:epsilon_distribution}c). 
A narrow band is evident around the edges of the nanoribbon that corresponds to a broadly distributed Gaussian with a center-of-mass significantly shifted from the spectrally-averaged mean determined by THz-SNOM (TDS-mode). 
The spatial distribution of this cluster, localized to the edges of the nanoribbon, agrees well with regions where we identify a high degree of local surface curvature (inferred from the second-derivative of the AFM topography, see~Fig.~\ref{fig:spatial_curvature}).
The curvature of TMD films determined in this way has previously been connected with areas of increased local strain~\cite{rahaman2017highly}. 
Tensile and compressive strain can influence the dielectric properties of materials. 
A compressive strain less than 10\% in monolayer MoS$_2$ has been predicted by calculations to result in a lowered static permittivity~\cite{kumar2013mechanical}; typically the reduction is on the order of several percent, which cannot fully account for the shift away from the mean observed for this edge cluster. 
We therefore suggest that other influences, including artifacts of the measurement from the sample topography and not from the material properties convolute the data in this region.    
The second cluster (with index 1) is larger, but shows a similarly shifted distribution in the real-part of the permittivity, well separated from the two final clusters (index 2,3) that comprise the spatial core of the nanoribbon and show the tightest distribution around the mean based on the nanoscopy data.  
Importantly, these core clusters of the nanoribbon are distinguishable by their permittivity distributions in the complex plane. 
We propose that this could arise due to subtle differences in their nanoscale material properties, including variations in local carrier concentration or the spatially integrated density of point defects.    

Finally, we note that although the dependence of the extracted effective complex permittivity is essentially independent of the MoS$_2$ nanoribbon thickness (see Fig.~\ref{fig:height_distribution}), a transition from a lower to a higher real-part permittivity appears to occur around 8-9~nm (corresponding to approximately 13 atomic layers). 
An increased surface-to-volume ratio and reduced dimensionality of materials can be associated with an reduction in both the dielectric permittivity and the scattering time of mobile charge carriers in semiconductors and metals, due to the breaking of polarizable bonds and enhanced scattering---surface effects modifying the optical and electrical properties of materials~\cite{Yoo2008DielectricConstantReduction}.   
This could indicate an evolution of the optical properties from single- and few-layer characteristics to a bulk-like behavior.  
However, we again note that the thinner regions are proximal to the edges of the nanoribbon and therefore may be more strongly dominated by edge effects, as discussed above and only weakly dependent on dimensionality.  

\section{Conclusion} 
We have quantitatively investigated the nanoscale optical properties of MoS$_2$ nanoribbons using THz-SNOM. 
Initial WL imaging of an isolated nanoribbon, together with satellite crystallites, revealed a clear spatially-resolved contrast in the scattered electric near-fields. 
THz nanoscopy along a cross section of the nanoribbon allowed us to extract the effective complex permittivity that showed a featureless spectrum within our probe bandwidth, well below the lowest frequency resonances of structural phonon modes in the material, indicating a low carrier concentration consistent with the approximately frequency independent behavior of the complex permittivity of a Drude conductor with a low scattering time.
By exploiting the phase information from the line scan measurements, where full scattered THz waveforms are detected in the time domain, we are able to extract spatially resolved maps of the complex permittivity of the nanoribbon from WL imaging data, with only a small error due to the rapid detection method being sensitive only to changes in the peak of the scattered THz waveform.  
The permittivity images reveal significant nanoscale variations in the optical properties of the nanoribbon (and its crystallites).
A clustering analysis allows us to resolve four dominant distributions that are separable in the complex plane and map them back into real-space.
This allows us to propose mechanisms, such as local strain gradients proximal to the nanoribbon edges or distributions in the spatially-averaged density of point defects that could be driving changes to the permittivity over such length-scales, and importantly move towards being able to unambiguously distinguish heterogeneities in nanoscale material properties from measurement artifacts, such as topographic effects, known to challenge interpretation of scanning probe techniques, including THz-SNOM.          
With very few methodologies available for either the direct quantification or indirect evaluation of the complex optical and electrical properties of materials at the nanoscale and the importance of understanding dielectric behavior, including disorder, for the development of opto- and nanoelectronic devices based on layered materials, we believe this approach will be useful for studying many other semiconducting nanomaterial systems in the future.    

\section*{Funding}
S.C. and E.J.R.K acknowledge financial support from Independent Research Fund Denmark Sapere Aude grants (project number 8049-00095B and project number 9064-00072B).
\section*{Author contributions}
SC, PUJ and EJRK conceived the idea. 
HBL, PUJ and EJRK planned and designed the experiments. 
The sample, along with the SEM image and Raman spectrum, were provided by DIM, GG and SC.
HBL performed the experiments. 
HBL performed the data analysis with contributions from WVC, PUJ and EJRK. 
WVC performed the statistical cluster analysis. 
HBL wrote the manuscript with contributions from WVC and EJRK. 
Edits were provided by all authors. 
PUJ and EJRK supervised the entire project.
All authors have accepted responsibility for the entire content of this manuscript and approved its submission.
\section*{Data availability statement}
Data sets generated and / or analyzed during the current study are available from the corresponding author on reasonable request.

\onecolumngrid
\setcounter{section}{0}
\setcounter{equation}{0}
\setcounter{figure}{0}
\renewcommand{\theequation}{S\arabic{equation}}
\renewcommand{\thefigure}{S\arabic{figure}}
\renewcommand{\thesubsection}{S\arabic{subsection}}
\section*{Supplementary Information for Dielectric permittivity extraction of MoS$_2$ nanoribbons using THz nanoscopy}
\subsection{FDM model and parameters}

Simulating the system's response through the FDM requires assumptions about the physical parameters used in the model, such as the tip radius $R_t$, the effective length of the extended dipole $L$, the experimentally determined fill factor $g$, and the in-plane momentum $q$.
This section will discuss the parameter selection used for modelling the THz-SNOM response and ultimately extract the physical sample parameters.

The FDM predicts the scattering contrast, $\eta_i$, by comparing the $i$-th demodulated scattered electric fields from the sample and substrate, respectively.
The contrast is directly proportional to the effective polarizability of the tip, such that:
\begin{equation}
\eta_i=\frac{E_{i}^{sample}}{E_{i}^{ref}}\propto\frac{\alpha_i^{sample}}{\alpha_i^{ref}}.
\end{equation}
The effective polarizability, $\alpha$, is dependent on the geometry of the system ($f_{0,1}$) and the properties of the sample ($\beta$) as shown in the main text Eq.~\ref{eq:effective_polarizability}.
The geometry parameters $f_{0,1}$ are dependent on the tip size, shape, and position above the sample.
The height dependence makes the geometry factor time-dependent.
The near-field reflection coefficient, $\beta$, is derived using the transfer matrix method of Zhan~{\em et al.}~\cite{Zhan2013}.
The tip apex radius is estimated from scanning electron microscopy (SEM) images of the standard tips used (Rocky Mountain Nanotechnology 25PtIr200B-40H), with an average of approximately 30~nm. 
The manufacturer pre-defines the shank length with a total length of $80~\mathrm{\mu}$m.
The in-plane momentum is determined as the maximum of the weighted distribution of the in-plane momentum distribution given by~\cite{Hauer2012, Wang2003_CHWOO}
\begin{equation}
    W(A, R_t, q) = q e^{-2qR_t}e^{q\left(H+R\right)},
\end{equation}
using a tip radius of 30 nm, yields a maximum of the weighted distribution at $q = 4.99\cdot 10^{7} \mathrm{m}^{-1}$.

Mooshammer~{\em et al.} showed through spatial representation of the demodulation orders that the effective width of the probe volume decreases with higher orders~\cite{doi:10.1021/acsphotonics.9b01533}. 
The effective in-plane momentum is corrected for by their findings by scaling with the relative change in probing width between the orders.
Finally, the permittivity of the substrate, which is selected based on the THz-TDS results of the refractive index of sapphire~\cite{Grischkowsky1990} as discussed in the main text, together with all other model parameters, is summarized in Table~\ref{tab:parameter_table}.

\begin{table}[htbp]
    \centering
    \begin{tabular}{c c c}
        Description             &  Parameter                & Value \\\hline
        Tip Radius              &      $R_\mathrm{t}$                    &  $30\:\mathrm{nm}$                          \\ 
        Dipole half-length      &      $L$                    &  $40\:\mathrm{\mu m}$                       \\
        In-plane momentum       &      $q$                    &  $2.47\cdot10^5 \:\mathrm{cm^{-1}}$         \\
        Tapping amplitude       &      $A$                    &  $170\:\mathrm{nm}$                         \\
        Substrate permittivity  & $\mathrm{\epsilon_{\perp}}$ &  $11.68$                                    \\ 
        Substrate permittivity  & $\mathrm{\epsilon_{\parallel}}$ &  $9.46$                                    \\ 
        Angle of incidence      &       $\mathrm{\theta}$   &  $30^{\circ}$                                          \\
    \end{tabular}
    \caption{Parameters used in the FDM model for permittivity extraction.}
    \label{tab:parameter_table}
\end{table}

\subsection{Permittivity extraction method}
Inversion of THz-SNOM data to permittivity values was performed numerically using a minimization algorithm.
The goal of the algorithm was to find the permittivity parameters ($\epsilon_r,\:\epsilon_i$) that resulted in the best overlap between simulation and measurement.
The algorithm uses the differential evolution method from SciPy.
The error function used for minimization is an expanded version of the one applied by Ritchie {\em et al.}~\cite{Ritchie2022} in combination with the iterative method of Mooshammer {\em et al.}~\cite{mooshammer2018nanoscale}.
The iterative part was dynamically varying permittivity parameters to find the lowest value of the error function and feed it back into the system.
Our extension of the Ritchie {\em et al.} error function consists of two additional terms that allow the algorithm to minimize multiple orders simultaneously. 
The original two error terms are given as
\begin{equation}
    E_{abs,i} = \frac{S_{exp,i} - S_{FDM,i}}{0.5(S_{exp,i} + S_{FDM,i})},\quad     E_{phase,i} = \frac{\phi_{exp,i} - \phi_{FDM,i}}{0.5(\phi_{exp,i} + \phi_{FDM,i})},
    \label{eq:original_error_function}
\end{equation}
where $S_i$ represents the absolute value and $\phi_i$ represents the angle of the scattering contrast of the $i$-th order. 
When applying the algorithm to multiple orders using only the terms in Eq.~\ref{eq:original_error_function}, it will not minimize for all orders simultaneously, but rather the absolute value of a single order and the phase of a single order. 
To mitigate this behavior, a penalty is introduced for optimizing better for some orders than others.
The penalty is the standard deviation of the errors in the absolute and phase values, respectively
\begin{equation}
    E_{penalty} = STD(E_{abs}) + STD(E_{phase}) = \sigma_{abs} + \sigma_{phase}.
\end{equation}
The total error function can then be described as
\begin{equation}
    E_{tot} = \sum \left(E_{abs, i} + E_{phase, i}\right) + \sigma_{abs} + \sigma_{phase}.
\end{equation}
This method is applied to each frequency individually and to each pixel individually in the WL images.

When inverting the WL imaging, each pixel value not only represents the absolute value of the frequency average, but is a mix of the phase and amplitude response.
This is due to the WL measurement using a fixed time delay for all spatial positions in the scanned area.
A phase shift of the pulse moves its peak away from the reference time delay, effectively decreasing the amplitude.
As discussed in the main text, the frequency-dependent phase shift was assumed constant, based on inspection of the THz nanoscopy data.
By taking the full THz waveform measured on the sample and applying the inverse phase shift, we can estimate the error in WL amplitude caused by a variable phase contribution.

\begin{figure}[htb]
    \centering
    \includegraphics[width=0.9\linewidth]{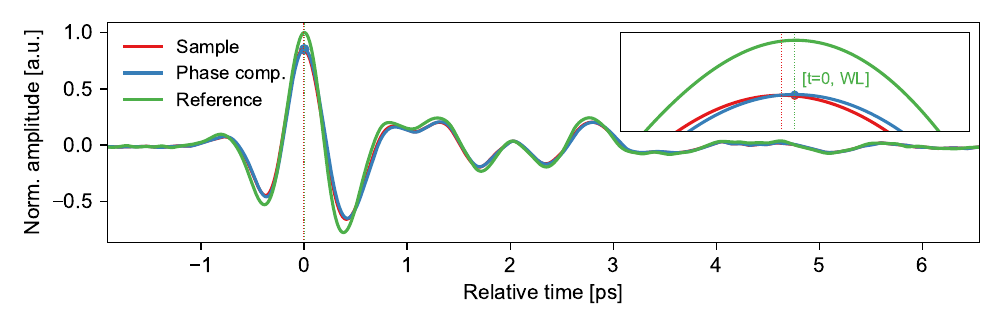}
    \caption{Second-order demodulated waveforms recorded on the substrate (green) and the sample (red), shown with a phase-corrected sample measurement (blue). The waveform is normalized to the peak of the reference waveform. The inset shows a magnified view of the peak of the waveforms with the vertical lines indicating the peak positions.}
    \label{fig:si_waveforms}
\end{figure}

Figure \ref{fig:si_waveforms} shows the waveforms of the second-order demodulated signal measured on the sample (red) and on the substrate (green) which is used in the main analysis.
The average phase shift is applied inversely on the sample measurement recovering the blue signal, which has better temporal overlap with the substrate at the peak.
The amplitude difference at the peak between the red and the blue curves is approximately $0.5\%$.

\subsection{Permittivity comparison from different harmonic orders}
When extracting the permittivity through the minimization routine described above, each harmonic should ideally return the same value, assuming there is no noise. 
In reality, the signal is noisy, resulting in each order, when used separately, returning slightly different values.

\begin{figure}[htbp]
    \centering
    \includegraphics[width=0.9\linewidth]{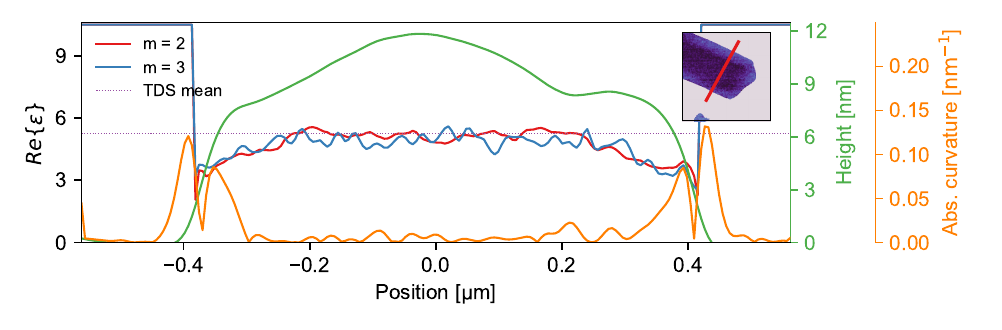}
    \caption{Line profile taken from the zoom-in permittivity map for the $i=2$, and $i=3$ harmonics. The position of the line profile is indicated by the red line on the inset. It should be noted, the extracted profile is not at the same position on the nanoribbon as where the THz nanoscopy was performed, as is discussed in the main text. The height from the substrate (or sample thickness) and curvature are shown in green and orange.}
    \label{fig:eps_comparison}
\end{figure}

The minimization performed on the orders separately for a cross-section across the zoom-in scan of the nanoribbon is shown in Fig.~\ref{fig:eps_comparison}. 
The central area of the nanoribbon shows a noisy plateau in permittivity, although the topography varies by several nanometers. 
This suggests that the material parameters are independent of height and that the method of extraction is robust. 
Using multiple orders will result in the same permittivity with less noise.

In Fig.~\ref{fig:eps_comparison}, a band of approximately $100~\mathrm{nm}$ around the nanoribbon can be seen with lower permittivity.
The orange curve shows the curvature of the nanoribbon, which can indicate a region of increased strain in the material~\cite{rahaman2017highly}, which in turn can affect the local permittivity of the material~\cite{YUE20121166}. 
The curvature falls off faster than the permittivity recovers when moving toward the center of the nanoribbon. 
This implies that the lowered permittivity cannot be fully explained by the impact of strain causing a change in the permittivity. 
Furthermore, previous reports have suggested nanoribbons are less prone to a build-up of residual strain, due to different strain-releasing processes during growth~\cite{Miakota2023MoS2NR}.

\begin{figure}[htbp]
    \centering
    \includegraphics[width=0.9\linewidth]{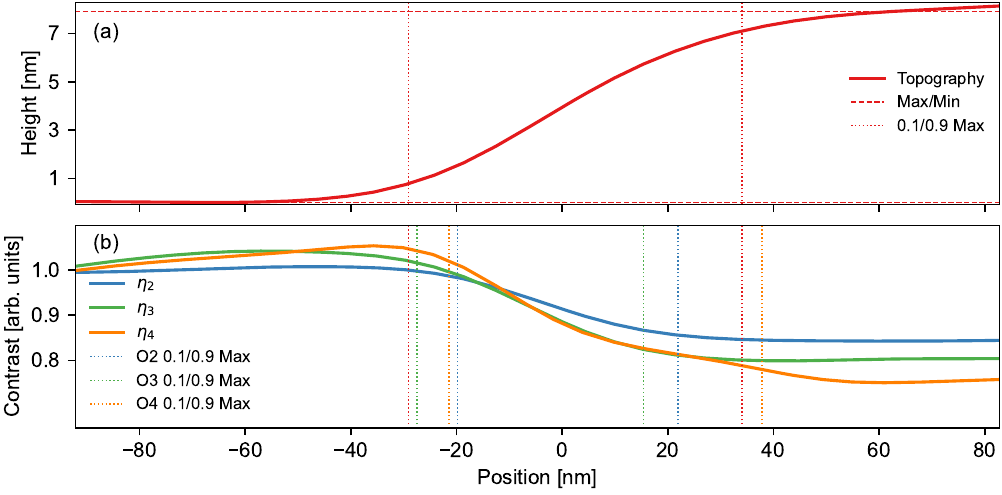}
    \caption{(a) AFM topography over an edge shown with the 10\% and 90\% positions of the lower and upper plateaus. (b) THz-SNOM WL contrast for orders 2-4 with their respective 10/90\% lines.}
    \label{fig:si_topographic_impact_on_optical}
\end{figure}

The FDM assumes an infinite and uniform layered stack below the tip. 
Proximal to edges several artefacts can occur, such as edge darkening, where the effective signal is reduced due to a rapid change in surface topography~\cite{Taubner2003}.
The lower signal strength would then result in an effectively lower permittivity.
However, the change in WL signal over an edge is wider than the step itself. 
This is illustrated in Fig.~\ref{fig:si_topographic_impact_on_optical}. 
Here the topography and WL signal for orders 2-4 are shown with the 10\% and 90\% positions of the step height. 
The width of the step is roughly $60$~nm, while the WL signal does not stabilise for another few tens of nanometers.
Each order is normalized to one in the substrate.
To the left of the edge, all values for the third and fourth orders are greater than one, suggesting that all the data immediately to the left of the nanoribbon is impacted by the nearby edge.

\subsection{Clustering}
Using a combination of WL data from the harmonic orders and the calculated permittivity values, it is possible to perform clustering to identify areas with similar properties. 
The clustering algorithm used in this work is a Bayesian Gaussian mixture model. 
This model determines multi-dimensional Gaussian distributions that fit the data and assigns the data to the cluster with the highest probability based on the discovered functions.
Based on calculations of the Akaike information criterion using different numbers of clusters, the optimal number of clusters was determined to be four. 
This number of clusters has a low prediction error while avoiding overfitting the data.

\begin{figure}[htbp]
    \centering
    \includegraphics[width=0.9\linewidth]{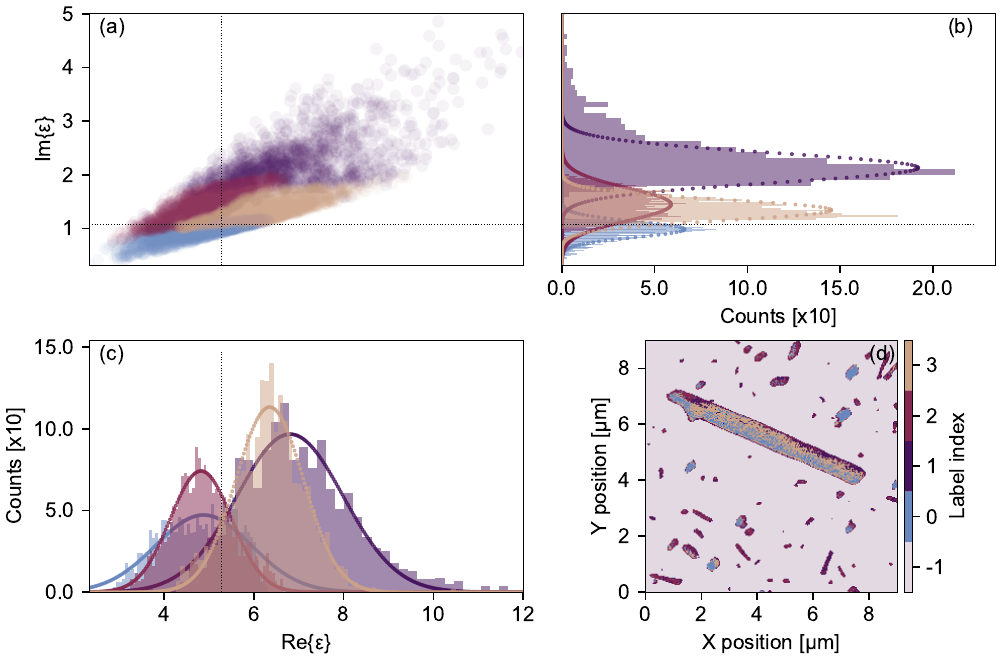}
    \caption{Clustering analysis of the spatially resolved permittivity data in the WL overview. (a) Four distinct clusters in the complex plane. Projections of the four dominant cluster distributions onto the imaginary (b) and real (c) axis. (d) Mapping of the clusters back to real-space to identify variations in the nanoscale permittivity associated with local changes in material properties.}
    \label{fig:epsilon_distribution_overview}
\end{figure}

The results of the algorithm on the zoom-in area of the nanoribbon are shown in the main text. 
Here, the results for the nanoribbon overview are presented (see Fig.~\ref{fig:epsilon_distribution_overview}).
This map captures the satellite MoS$_2$ crystallites. 
The close-up scan showed distributions indicating areas where topography and material properties are convoluted, as well as areas with consistent material properties. 
The overview scan reveals many areas with potentially varying properties.

The clustering for the overview cannot separate the topographically entangled points from the material properties due to the large proportion of these points, lower resolution, and more total edge points. 
In this case, the clusters cannot be interpreted as areas with constant properties and Gaussian noise but rather as areas with similar results from the entanglement of material properties and topographic effects.
However, certain crystallites appear to belong to one cluster (index 0), while others belong to another cluster (index 1). 
The reason for this is worthly of further investigation, but could indicate a difference in growth conditions, residual local strain, or topography-induced artefacts. 
Ultimately the signal quality and spatial resolution of the scan limits the spatial separation of the clusters.

\begin{figure}[!t]
    \centering
    \includegraphics[width=0.9\linewidth]{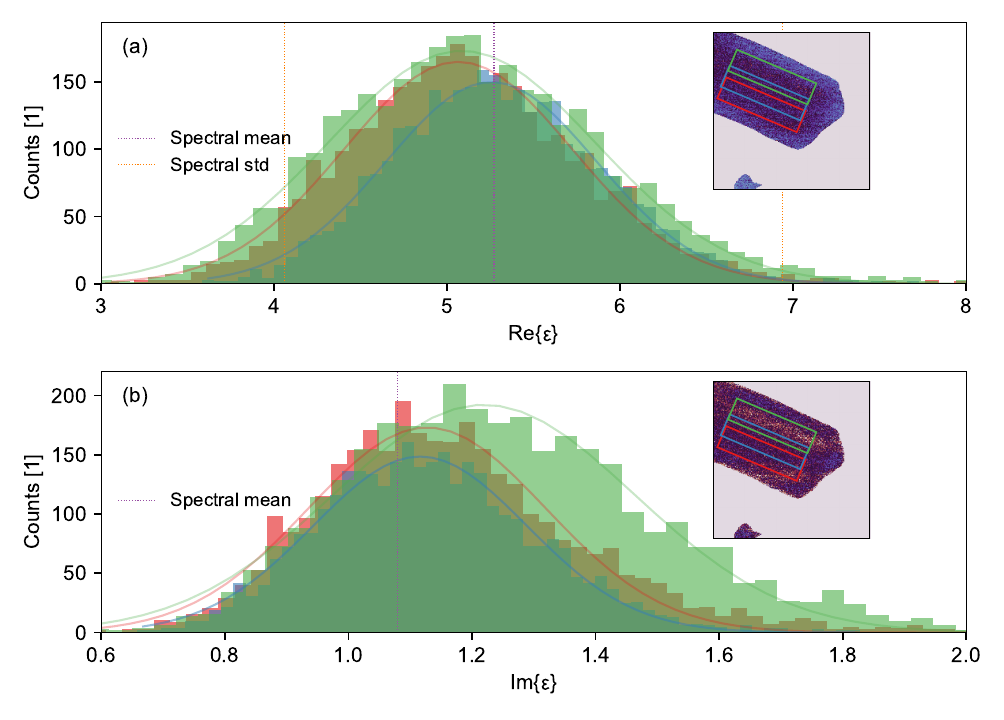}
    \caption{Real (a) and imaginary (b) permittivity distributions extracted from three overlapping areas (colour-coded to the correspondign distribution) from within the core of the nanoribbon, based on inversion of the WL image of the zoom-in (as indicated in the insets, respectively).}
    \label{fig:cutout_histograms}
\end{figure}
By making cut-outs from within the nanoribbon core it is possible to analyse nanoscale changes in the permittivity without considering edge artefacts or the influence of strain (due to a relatively constant profile of the curvature throughout the selected regions).
The core of the nanoribbon is segmented into three sections as shown in Fig.~\ref{fig:cutout_histograms}.
The corresponding distribution of the real and imaginary parts of the extracted permittivity from each section is plotted and fitted with a single Gaussian function.
Each region shows a clearly separated distribution suggesting indeed we capture variations in the effective complex permittivity on sub-micrometer length-scales that could arise due to local changes in carrier or defect concentration.
The vertical lines indicate the mean and standard deviation taken from the spectrally inverted data based on a full-field analysis of the scattered THz waveform.

\subsection{Curvature and strain}
\begin{figure}[!t]
    \centering
    \includegraphics[width=0.9\linewidth]{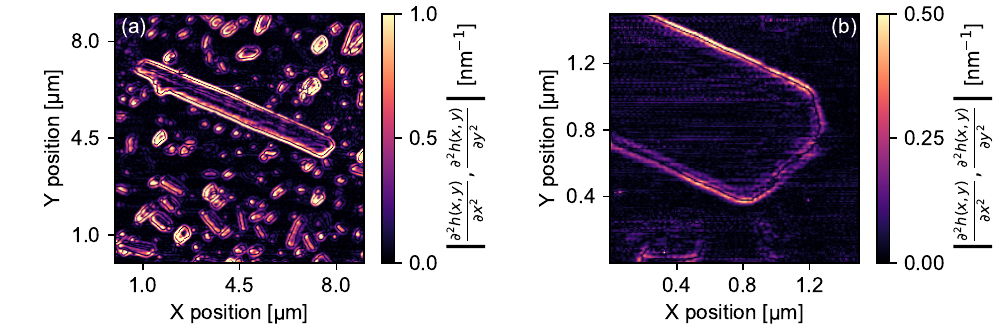}
    \caption{Spatial maps of the absolute value of the second derivative of the topography for the overview (a) and the zoom-in (b).}
    \label{fig:spatial_curvature}
\end{figure}
According to Rahaman~{\em et al.}, the curvature of the topographic map can be related to the magnitude of the local strain in a material~\cite{rahaman2017highly}. 
By applying the Laplacian to the topographic maps (both the overview and zoom-in) the curvature is extracted and shown in Fig.~\ref{fig:spatial_curvature}.
The curvature is largest near the edges and almost constant and zero across the core of the nanoribbon, this is in agreement with predictions of strain release in nanoribbons during the growth process. 
The curvature varies between the satellite crystallites suggesting possible differences in local strain induced during growth.

Yue et al. have shown that strain in $\mathrm{MoS_2}$ changes its permittivity depending on the magnitude and direction of the strain~\cite{YUE20121166}. 
This means the permittivity of the $\mathrm{MoS_2}$ nanoribbon might change near the edges. 
However, due to the impact of topographic cross-talk affecting the scattered near-field signal, further clarifying investigations are needed to separate the relative contributions of these various contributions to the observed spatial variations in permittivity. 

\subsection{Thickness dependence}
\begin{figure}[htbp]
    \centering
    \includegraphics[width=0.9\linewidth]{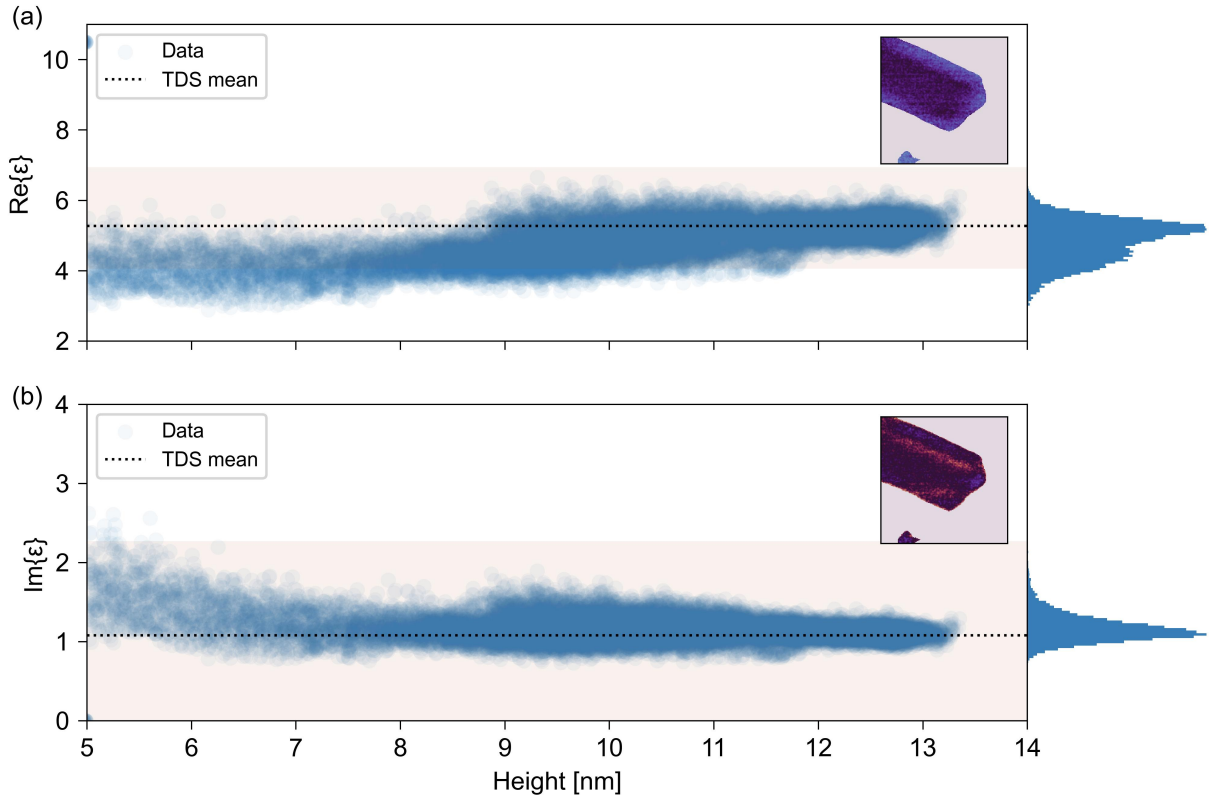}
    \caption{Thickness dependence of the real (a) and imaginary (b) parts of the permittivity inverted from the zoom-in WL imaging of the nanoribbon. Projections to the right show the corresponding histogram distributions. The horizontal dotted black line indicates the mean value from THz nanoscopy measurements and the corresponding standard deviation (transparent band).}
    \label{fig:height_distribution}
\end{figure}
We show how the real and imaginary parts of the extracted permittivity dependent insensitively on the sample thickness (see Fig.~\ref{fig:height_distribution}a and b). 
A small, but noticeable shoulder around 9~nm in the real-part could indicate a transition to bulk-like behaviour; however, the real-space image suggests the regions with reduced thickness correlated with a reduced real-part permittivity comprise a band proximal to the nanoribbon edge. 
This once again indicates that cross-talk due to topographic artefacts as the probe transitions between the nanoribbon and substrate make it difficult to draw a strong conclusion without further clarifying investigations.


\begin{thebibliography}{99}

\bibitem{raja2019dielectric}
A.~Raja, {\em et~al.}, ``Dielectric disorder in two-dimensional materials,'' {\em Nat. Nanotechnol.}, vol.~14, no.~9, pp.~832--837, 2019.

\bibitem{Varghese2020}
A.~Varghese, {\em et~al.}, ``Near-direct bandgap {W}{S}e$_2$/{R}e{S}$_2$ type-II pn heterojunction for enhanced ultrafast photodetection and high-performance photovoltaics,'' {\em Nano Lett.}, vol.~20, no.~3, pp.~1707--1717, 2020.

\bibitem{Furchi2014}
M.~M. Furchi, A.~Pospischil, F.~Libisch, J.~Burgd{\"o}, and T.~Mueller, ``Photovoltaic effect in an electrically tunable Van der Waals heterojunction,'' {\em Nano Lett.}, vol.~14, no.~8, pp.~4785--4791, 2014.

\bibitem{HuJAP2014}
T.~Hu, J.~Zhou, J.~Dong, and Y.~Kawazoe, ``Electronic and magnetic properties of armchair MoS$_2$ nanoribbons under both external strain and electric field, studied by first principles calculations,'' {\em J. Appl. Phys.}, vol.~116, no.~6, pp.~064301, 2014.

\bibitem{LiNaFNR2018}
S.~Li, {\em et~al.}, ``Vapour–liquid–solid growth of monolayer MoS$_2$ nanoribbons,'' {\em Nat. Mater.}, vol.~17, no.~6, pp.~535--542, 2018.

\bibitem{LiMos2NR2018}
Y.~Li, {\em et~al.}, ``Large-scale fabrication of MoS$_2$ ribbons and their light-induced electronic/thermal properties: Dichotomies in the structural and defect engineering,'' {\em Adv. Funct. Mater.}, vol.~28, no.~13, pp.~1704863, 2018.

\bibitem{Chowdhury2020}
T.~Chowdhury, {\em et~al.}, ``Substrate-directed synthesis of {M}o{S}$_2$ nanocrystals with tunable dimensionality and optical properties,'' {\em Nat. Nanotechnol.}, vol.~15, no.~1, pp.~29--34, 2020.

\bibitem{Ghimire2023MoS2NR}
G.~Ghimire, {\em et~al.}, ``Molybdenum disulfide nanoribbons with enhanced edge nonlinear response and photoresponsivity,'' {\em Adv. Mater.}, vol.~35, no.~31, pp.~2302469, 2023.

\bibitem{Zhang2021}
X.~Zhang, {\em et~al.}, ``Controllable epitaxial growth of large-area MoS$_2$/WS$_2$ vertical heterostructures by confined-space chemical vapor deposition,'' {\em Small}, vol.~17, no.~18, pp.~2007312, 2021.

\bibitem{NatCom2015ALDMoS2}
J.-G. Song, {\em et~al.}, ``Controllable synthesis of molybdenum tungsten disulfide alloy for vertically composition-controlled multilayer,'' {\em Nat. Commun.}, vol.~6, no.~1, pp.~7817, 2015.

\bibitem{NLMoS2CVDgen2014}
D.~J. Clark, {\em et~al.}, ``Strong optical nonlinearity of cvd-grown {M}o{S}$_2$ monolayer as probed by wavelength-dependent second-harmonic generation,'' {\em Phys. Rev. B}, vol.~90, no.~12, pp.~121409, 2014.

\bibitem{SHG2015WSe2opticsNat}
K.~L. Seyler, {\em et~al.}, ``Electrical control of second-harmonic generation in a {W}{S}e$_2$ monolayer transistor,'' {\em Nat. Nanotechnol.}, vol.~10, no.~5, pp.~407--411, 2015.

\bibitem{Morozov2015OpticalConstantsDynamic}
Y.~V. Morozov and M.~Kuno, ``Optical constants and dynamic conductivities of single layer {{MoS$_2$}}, {{MoSe$_2$}}, and {{WSe$_2$}},'' {\em Appl. Phys. Lett.}, vol.~107, no.~8, pp.~083103, 2015.

\bibitem{7.5strain}
Z.~Li, {\em et~al.}, ``Efficient strain modulation of 2d materials via polymer encapsulation,'' {\em Nat. Commun.}, vol.~11, no.~1, 2020.

\bibitem{BandGapEng_2013}
H.~J. Conley, B.~Wang, J.~I. Ziegler, R.~F. Haglund, S.~T. Pantelides, K.~I. Bolotin, ``Bandgap engineering of strained monolayer and bilayer {M}o{S}${_2}$,'' {\em Nano Lett.}, vol.~13, no.~8, pp.~3626–3630, 2013.

\bibitem{Young2017}
J.~R. Young, {\em et~al.}, ``Uniform large-area growth of nanotemplated high-quality monolayer {M}o{S}${_2}$,'' {\em Appl. Phys. Lett.}, vol.~110, no.~26, pp.~263103, 2017.

\bibitem{Zhang2015}
H.~Zhang, {\em et~al.}, ``Measuring the refractive index of highly crystalline monolayer {M}o{S}${_2}$ with high confidence,'' {\em Sci. Rep.}, vol.~5, pp.~8440, 2015.

\bibitem{Ermolaev2020}
G.~A. Ermolaev, {\em et~al.}, ``Broadband optical properties of monolayer and bulk {M}o{S}${_2}$,'' {\em npj 2D Mater Appl}, vol.~4, pp.~21, 2020.

\bibitem{zhao2017probing}
P.~Zhao, {\em et~al.}, ``Probing interface defects in top-gated {M}o{S}${_2}$ transistors with impedance spectroscopy,'' {\em ACS Appl. Mater. Interfaces}, vol.~9, no.~28, pp.~24348--24356, 2017.

\bibitem{Jepsen2011}
P.~Jepsen, D.~Cooke, and M.~Koch, ``Terahertz spectroscopy and imaging -- {{Modern}} techniques and applications,'' {\em Laser Photonics Rev.}, vol.~5, no.~1, pp.~124--166, 2011.

\bibitem{buron2012graphene}
J.~D. Buron, {\em et~al.}, ``Graphene conductance uniformity mapping,'' {\em Nano Lett.}, vol.~12, no.~10, pp.~5074--5081, 2012.

\bibitem{Yan2015}
X.~Yan, L.~Zhu, Y.~Zhou, Y.~E, L.~Wang, and X.~Xu, ``Dielectric property of {M}o{S}${_2}$ crystal in terahertz and visible regions,'' {\em Appl. Opt.}, vol.~54, no.~22, pp.~6732--6736, 2015.

\bibitem{Hou2022QuantificationDielectricConstant}
Y.~Hou, G.~Wang, C.~Ma, Z.~Feng, Y.~Chen, and T.~Filleter, ``Quantification of the dielectric constant of {M}o{S}${_2}$ and {W}{S}e${_2}$ Nanosheets by electrostatic force microscopy,'' {\em Mater. Charact.}, vol.~193, pp.~112313, 2022.

\bibitem{Knoll2000}
B.~Knoll and F.~Keilmann, ``Enhanced dielectric contrast in scattering-type scanning near-field optical microscopy,'' {\em Opt. Commun.}, vol.~182, pp.~321--328, 2000.

\bibitem{Keilmann2004}
F.~Keilmann and R.~Hillenbrand, ``Near-field microscopy by elastic light scattering from a tip,'' {\em Philosophical Transactions of the Royal Society A: Mathematical, Physical and Engineering Sciences}, vol.~362, no.~1817, pp.~787--805, 2004.

\bibitem{Chen2019ModernScatteringTypeScanning}
X.~Chen, {\em et~al.}, ``Modern {{Scattering-Type Scanning Near-Field Optical Microscopy}} for {{Advanced Material Research}},'' {\em Adv. Mater.}, vol.~31, pp.~1804774, 2019.

\bibitem{Govyadinov2013}
A.~A. Govyadinov, I.~Amenabar, F.~Huth, P.~S. Carney, and R.~Hillenbrand, ``Quantitative measurement of local infrared absorption and dielectric function with tip-enhanced near-field microscopy,'' {\em J. Phys. Chem. Lett.}, vol.~4, no.~9, pp.~1526--1531, 2013.

\bibitem{Govyadinov2014}
A.~A. Govyadinov, S.~Mastel, F.~Golmar, A.~Chuvilin, P.~S. Carney, and R.~Hillenbrand, ``Recovery of permittivity and depth from near-field data as a step toward infrared nanotomography,'' {\em ACS Nano}, vol.~8, no.~7, pp.~6911–6921, 2014.

\bibitem{McLeod2014}
A.~S. McLeod, {\em et~al.}, ``Model for quantitative tip-enhanced spectroscopy and the extraction of nanoscale-resolved optical constants,'' {\em Phys. Rev. B}, vol.~90, no.~8, pp.~085136, 2014.

\bibitem{Ritchie2022}
E.~T. Ritchie, C.~B. Casper, T.~A. Lee, and J.~M. Atkin, ``Quantitative local conductivity imaging of semiconductors using near-field optical microscopy,'' {\em J. Phys. Chem. C}, vol.~126, no.~9, pp.~4515--4521, 2022.

\bibitem{zizlsperger2024situ}
M.~Zizlsperger, {\em et~al.}, ``In situ nanoscopy of single-grain nanomorphology and ultrafast carrier dynamics in metal halide perovskites,'' {\em Nat. Photonics}, vol.~18, no.~9, pp.~975--981, 2024.

\bibitem{mooshammer2018nanoscale}
F.~Mooshammer, {\em et~al.}, ``Nanoscale near-field tomography of surface states on ({B}i$_{0.5}${S}b$_{0.5}$)$_2${T}e$_2$,'' {\em Nano Lett.}, vol.~18, no.~12, pp.~7515--7523, 2018.

\bibitem{Cvitkovic2007}
A.~Cvitkovic, N.~Ocelic, and R.~Hillenbrand, ``Analytical model for quantitative prediction of material contrasts in scattering-type near-field optical microscopy,'' {\em Opt. Express}, vol.~15, pp.~8550--8565, 2007.

\bibitem{Hauer2012}
B.~Hauer, A.~P. Engelhardt, and T.~Taubner, ``Quasi-analytical model for scattering infrared near-field microscopy on layered systems,'' {\em Opt. Express}, vol.~20, pp.~13173--13188, 2012.

\bibitem{Wirth2021}
K.~G. Wirth, {\em et~al.}, ``Tunable s-snom for nanoscale infrared optical measurement of electronic properties of bilayer graphene,'' {\em ACS Photonics}, vol.~8, pp.~418--423, 2021.

\bibitem{Miakota2023MoS2NR}
D.~I. Miakota, G.~Ghimire, R.~{Kumar Ulaganathan}, M.~E. Rodriguez, and S.~Canulescu, ``A novel two-step route to unidirectional growth of multilayer {M}o{S}${_2}$ nanoribbons,'' {\em Appl. Surf. Sci.}, vol.~619, pp.~156748, 2023.

\bibitem{Ermolaev2021GiantOpticalAnisotropy}
G.~A. Ermolaev, {\em et~al.}, ``Giant optical anisotropy in transition metal dichalcogenides for next-generation photonics,'' {\em Nat. Commun.}, vol.~12, no.~1, pp.~854, 2021.

\bibitem{AnomalMoS2vibrLee2010}
C.~Lee, H.~Yan, L.~E. Brus, T.~F. Heinz, J.~Hone, and S.~Ryu, ``Anomalous lattice vibrations of single- and few-layer {M}o{S}${_2}$,'' {\em ACS Nano}, vol.~4, no.~5, pp.~2695--2700, 2010.

\bibitem{rahaman2017highly}
M.~Rahaman, {\em et~al.}, ``Highly localized strain in a {M}o{S}${_2}$/{A}u heterostructure revealed by tip-enhanced Raman spectroscopy,'' {\em Nano Lett.}, vol.~17, no.~10, pp.~6027--6033, 2017.

\bibitem{jing2023phase}
R.~Jing, {\em et~al.}, ``Phase-resolved terahertz nanoimaging of {W}{T}e$_2$ microcrystals,'' {\em Phys. Rev. B}, vol.~107, no.~15, pp.~155413, 2023.

\bibitem{Neu2018}
J.~Neu and C.~A. Schmuttenmaer, ``Tutorial: {{An}} introduction to terahertz time domain spectroscopy ({{THz-TDS}}),'' {\em J. Appl. Phys.}, vol.~124, pp.~231101, 2018.

\bibitem{schneider2005scattering}
S.~Schneider, S.~Grafstr{\"o}m, and L.~Eng, ``Scattering near-field optical microscopy of optically anisotropic systems,'' {\em Phys. Rev. B}, vol.~71, no.~11, pp.~115418, 2005.

\bibitem{Yao2021ProbingSubwavelengthInplane}
Z.~Yao, {\em et~al.}, ``Probing subwavelength in-plane anisotropy with antenna-assisted infrared nano-spectroscopy,'' {\em Nat. Commun.}, vol.~12, pp.~2649, 2021.

\bibitem{ruta2020quantitative}
F.~L. Ruta, A.~J. Sternbach, A.~B. Dieng, A.~S. McLeod, and D.~Basov, ``Quantitative nanoinfrared spectroscopy of anisotropic van der waals materials,'' {\em Nano Lett.}, vol.~20, no.~11, pp.~7933--7940, 2020.

\bibitem{norgaard2024near}
M.~N{\o}rgaard, T.~Yezekyan, S.~Rolfs, C.~Frydendahl, N.~A. Mortensen, and V.~A. Zenin, ``Near-field refractometry of van der waals crystals,'' {\em arXiv preprint arXiv:2411.07926}, 2024.

\bibitem{yao2021probing}
Z.~Yao, {\em et~al.}, ``Probing subwavelength in-plane anisotropy with antenna-assisted infrared nano-spectroscopy,'' {\em Nat. Commun.}, vol.~12, no.~1, pp.~2649, 2021.

\bibitem{Zhan2013}
T.~Zhan, X.~Shi, Y.~Dai, X.~Liu, and J.~Zi, ``Transfer matrix method for optics in graphene layers,'' {\em J. Phys.: Condens. Matter}, vol.~25, pp.~215301, 2013.

\bibitem{chen2021rapid}
X.~Chen, Z.~Yao, S.~G. Stanciu, D.~Basov, R.~Hillenbrand, and M.~Liu, ``Rapid simulations of hyperspectral near-field images of three-dimensional heterogeneous surfaces,'' {\em Opt. Express}, vol.~29, no.~24, pp.~39648--39668, 2021.

\bibitem{mastel2018understanding}
S.~Mastel, A.~A. Govyadinov, C.~Maissen, A.~Chuvilin, A.~Berger, and R.~Hillenbrand, ``Understanding the image contrast of material boundaries in IR nanoscopy reaching 5 nm spatial resolution,'' {\em ACS Photonics}, vol.~5, no.~8, pp.~3372--3378, 2018.

\bibitem{Grischkowsky1990}
D.~Grischkowsky, S.~Keiding, M.~V. Exter, and C.~Fattinger, ``Far-infrared time-domain spectroscopy with terahertz beams of dielectrics and semiconductors,'' {\em J. Opt. Soc. Am. B}, vol.~7, pp.~2006--2015, 1990.

\bibitem{mester2022high}
L.~Mester, A.~A. Govyadinov, and R.~Hillenbrand, ``High-fidelity nano-FTIR spectroscopy by on-pixel normalization of signal harmonics,'' {\em Nanophotonics}, vol.~11, no.~2, pp.~377--390, 2022.

\bibitem{molina2011phonons}
A.~Molina-Sanchez and L.~Wirtz, ``Phonons in single-layer and few-layer {M}o{S}$_2$ and {W}{S}$_2$,'' {\em Phys. Rev. B}, vol.~84, no.~15, pp.~155413, 2011.

\bibitem{kumar2013mechanical}
A.~Kumar and P.~Ahluwalia, ``Mechanical strain dependent electronic and dielectric properties of two-dimensional honeycomb structures of {M}o{X}$_2$ (X={S}, {S}e, {T}e),'' {\em Physica B}, vol.~419, pp.~66--75, 2013.

\bibitem{Yoo2008DielectricConstantReduction}
H.~G. Yoo and P.~M. Fauchet, ``Dielectric constant reduction in silicon nanostructures,'' {\em Phys. Rev. B}, vol.~77, np.~11, p.~115355, 2008.

\end{thebibliography}

\begin{thebibliography}{99}

\bibitem{Zhan2013}
T. Zhan, X. Shi, Y. Dai, X. Liu, and J. Zi, “Transfer matrix method for optics in graphene layers,” {\em J. Phys.: Condens. Matter}, vol.~25, p.~215301, 2013

\bibitem{Hauer2012}
B. Hauer, A. P. Engelhardt, and T. Taubner, “Quasi-analytical model for scattering infrared near-field microscopy on layered systems,” {\em Opt Express}, vol.~20, pp.~13173–13188, 2012.

\bibitem{Wang2003_CHWOO}
B. Wang and C. H. Woo, “Atomic force microscopy-induced electric field in ferroelectric thin films,” {\em J. Appl. Phys.}, vol.~94, no.~6, pp.~4053–4059, 2003.

\bibitem{doi:10.1021/acsphotonics.9b01533}
F. Mooshammer, M. A. Huber, F. Sandner, M. Plankl, M. Zizlsperger, and R. Huber, “Quantifying nanoscale electro-magnetic fields in near-field microscopy by fourier demodulation analysis,” {\em ACS Photonics}, vol.~7, no.~2, pp.~344–351, 2020.

\bibitem{Grischkowsky1990}
D. Grischkowsky, S. Keiding, M. V. Exter, and C. Fattinger, “Far-infrared time-domain spectroscopy with terahertz beams of dielectrics and semiconductors,” {\em J. Opt. Soc. Am. B}, vol.~7, pp.~2006–2015, 1990.

\bibitem{Ritchie2022}
E. T. Ritchie, C. B. Casper, T. A. Lee, and J. M. Atkin, “Quantitative local conductivity imaging of semiconductors using near-field optical microscopy,” {\em J. Phys. Chem. C}, vol.~126, pp.~4515–4521, 2022.

\bibitem{mooshammer2018nanoscale}
F. Mooshammer, {\em et~al.}, “Nanoscale near-field tomography of surface states on (Bi0.5Sb0.5)2Te3,” {\em Nano Lett.}, vol.~18, no.~12, pp.~7515–7523, 2018.

\bibitem{rahaman2017highly}
M. Rahaman, R. D. Rodriguez, G. Plechinger, S. Moras, C. Schüller, T. Korn, and D. R. Zahn, “Highly localized strain in a MoS2/Au heterostructure revealed by tip-enhanced raman spectroscopy,” {\em Nano Lett.}, vol.~17, no.~10, pp.~6027–6033, 2017.

\bibitem{YUE20121166}
Q. Yue, {\em et~al.}, “Mechanical and electronic properties of monolayer MoS2 under elastic strain,” {\em Phys. Lett. A}, vol.~376, no.~12, pp.~1166–1170, 2012.

\bibitem{Miakota2023MoS2NR}
D. I. Miakota, G. Ghimire, R. Kumar Ulaganathan, M. E. Rodriguez, and S. Canulescu, “A novel two-step route to unidirectional growth of multilayer MoS2 nanoribbons,” {\em Appl. Surf. Sci.}, vol.~619, p.~156748, 2023.

\bibitem{Taubner2003}
T. Taubner, R. Hillenbrand, and F. Keilmann, “Performance of visible and mid-infrared scattering-type near-field optical microscopes,” {\em J. Microsc.}, vol.~210, no.~3, pp.~311–314, 2003.

\end{thebibliography}
\end{document}